\documentclass[a4paper,11pt]{article}
\usepackage{pos}
\usepackage{comment}
\title{Fast Neutrino Flavor Conversion at Late Time}

\author*[a]{Soumya Bhattacharyya}
\affiliation[a]{Tata Institute of Fundamental Research, Homi Bhabha Road, Mumbai 400005, India}
\emailAdd{soumyaastroparticle@gmail.com}
\abstract{ We all know that in the dense anisotropic interior of the star, neutrino-neutrino forward-scattering can lead to fast collective neutrino oscillations, which has striking consequences on flavor dependent neutrino emission and can be crucial for the evolution of a supernova and its neutrino signal. The flavor evolution of such dense neutrino system is governed
	by a large number of coupled nonlinear partial differential equations which are almost always very difficult to solve. Although the triggering, initial linear growth and the condition  for fast oscillations to occur are understood by a well known trick known as ``Linear stability analysis'' \cite{Banerjee:2011fj}, this fails to answer an important question -- what is the impact
	of fast flavor conversion on observable neutrino fluxes or
	the supernova explosion mechanism? This is a significantly harder problem that requires understanding the nature of the final state solution in the nonlinear regime. Moving towards this direction we present one of the first numerical as well as an analytical study of the coupled flavor evolution of a non-stationary and inhomogeneous dense neutrino system in the nonlinear regime considering one spatial dimension and a spectrum of velocity modes. This work gives a clear picture of the final state flavor dynamics of such systems specifying its dependence on space-time coordinates, phase space variables as well as the lepton asymmetry and thus can have significant implications for the supernova astrophysics as well as its associated neutrino phenomenology even for the most realistic scenario.}

\FullConference{%
  40th International Conference on High Energy physics - ICHEP2020\\
  July 28 - August 6, 2020\\
  Prague, Czech Republic (virtual meeting)
}

\begin{document}
\maketitle	
\begin{figure*}[]
	\hspace{-1 cm}
	\includegraphics[width=0.24\columnwidth]{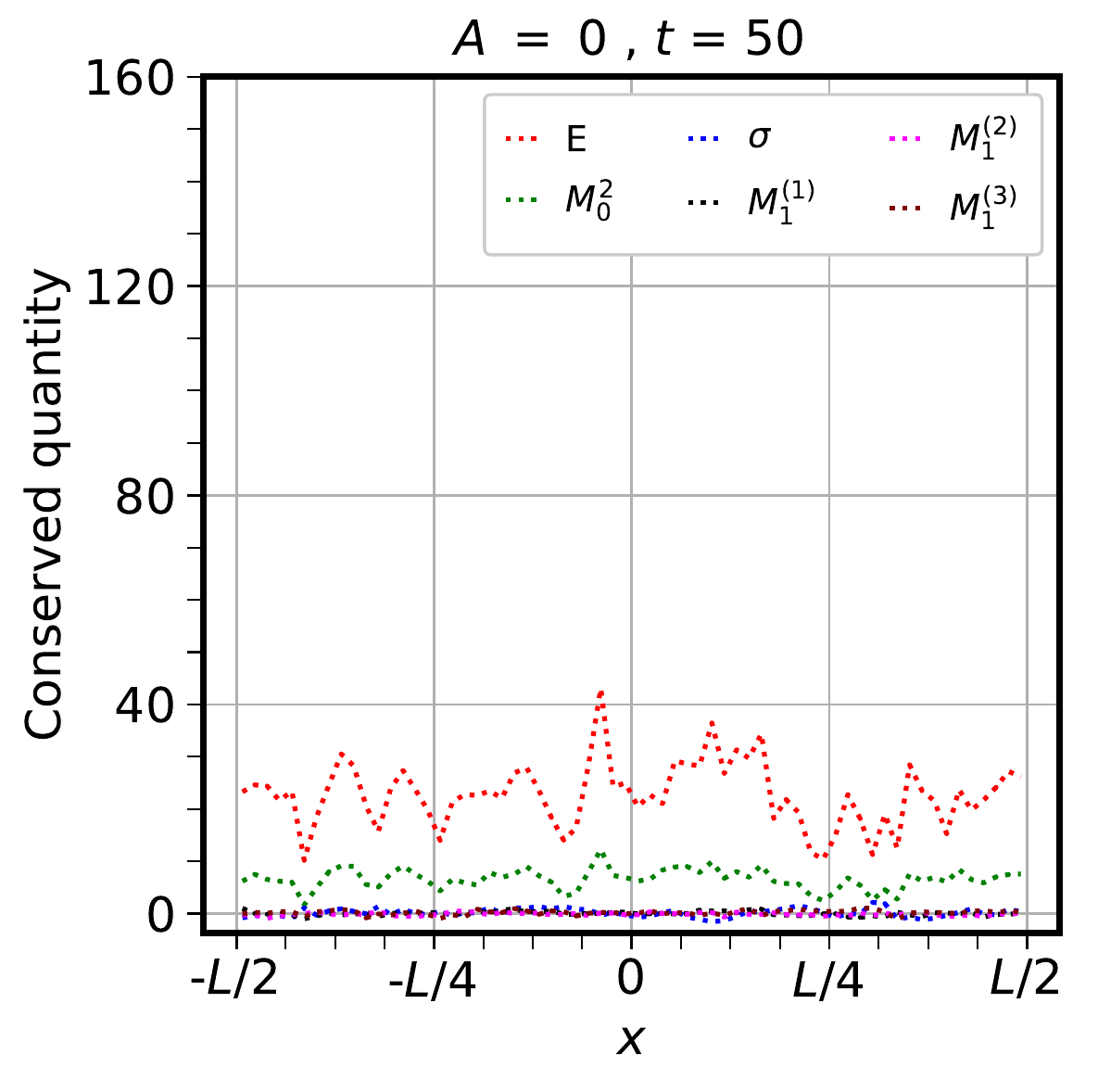}~
	\includegraphics[width=0.24\columnwidth]{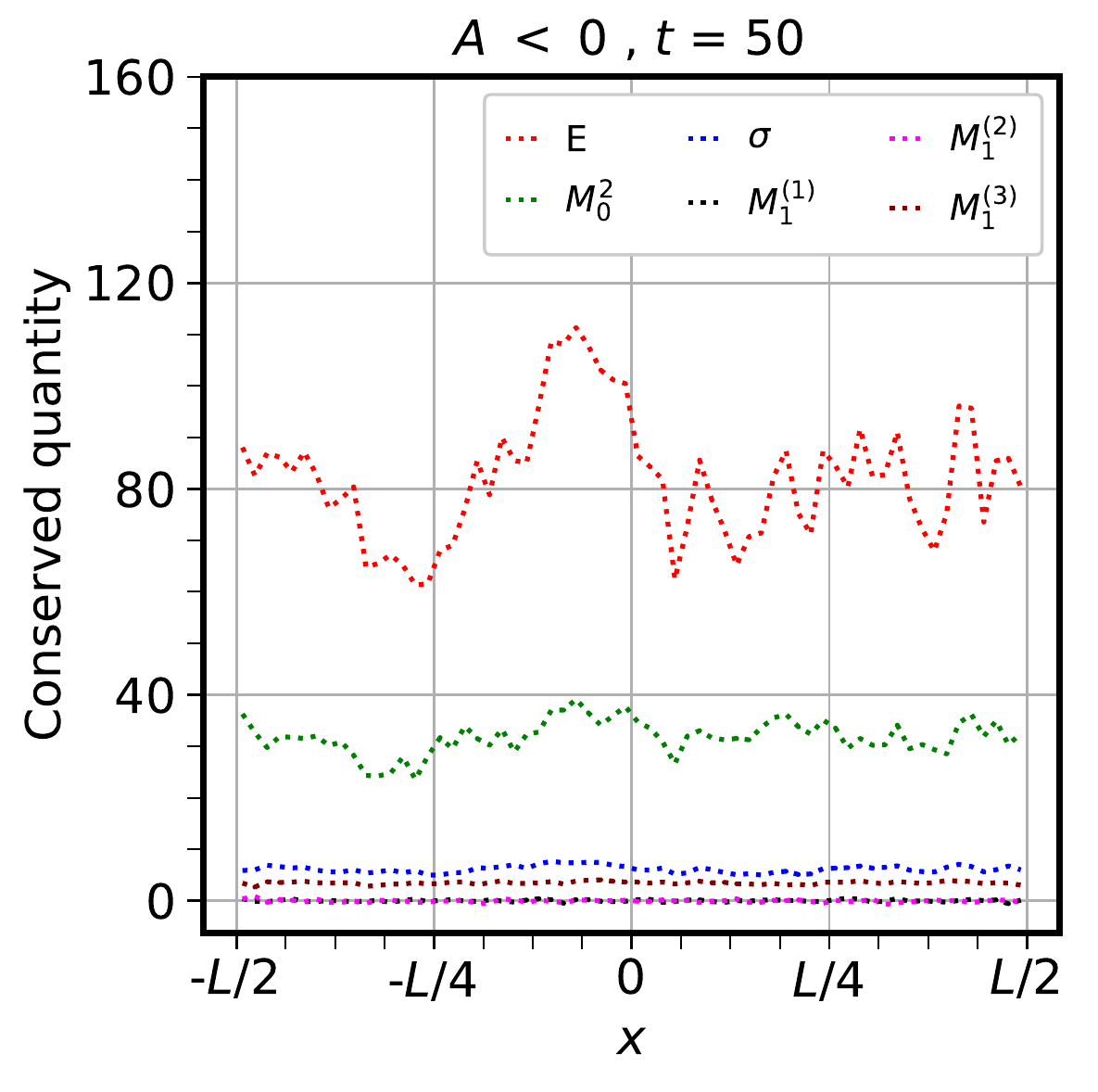}~
	\hspace{1.2 cm}
	\includegraphics[width=0.24\columnwidth]{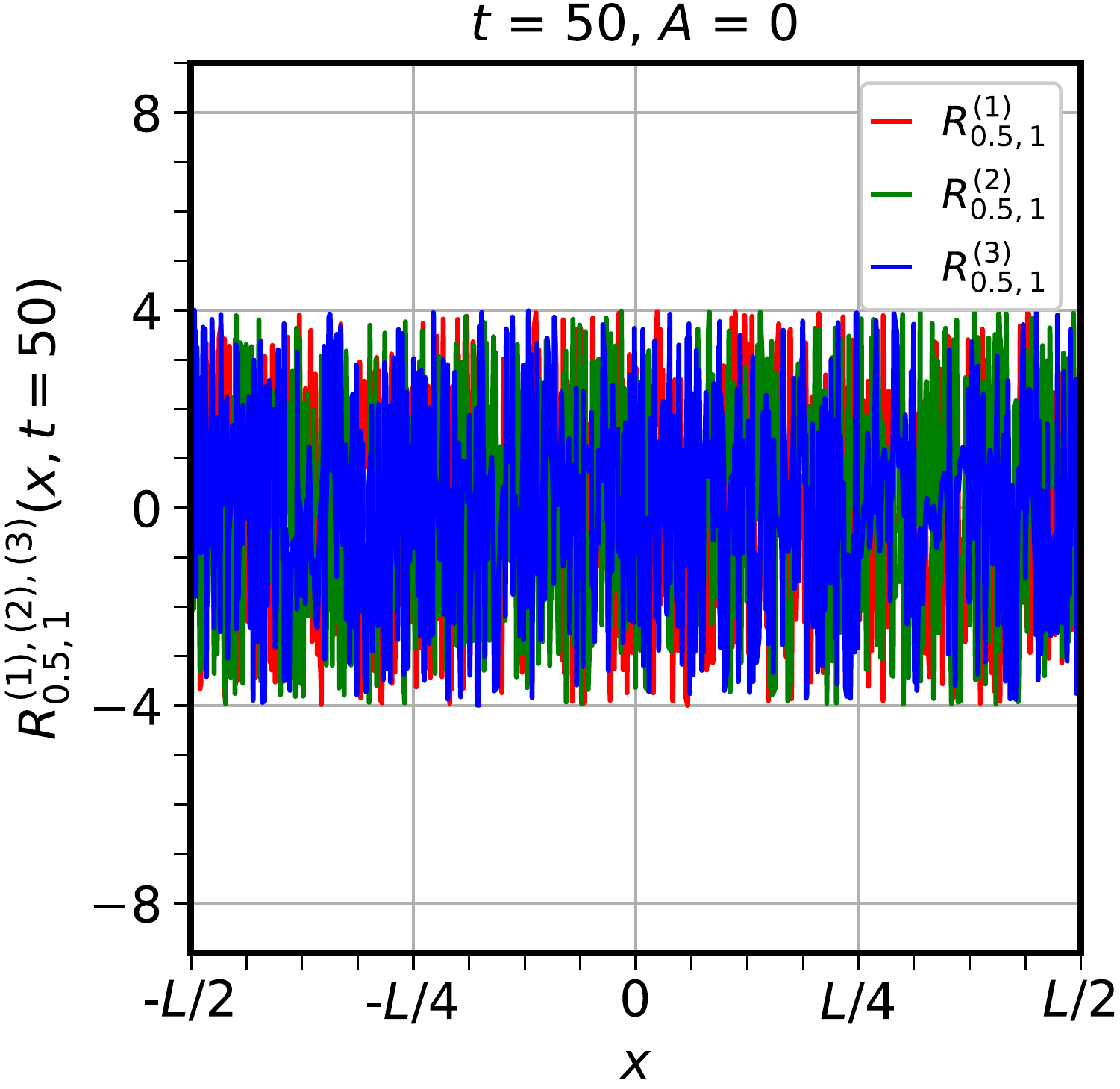}
	\includegraphics[width=0.24\columnwidth]{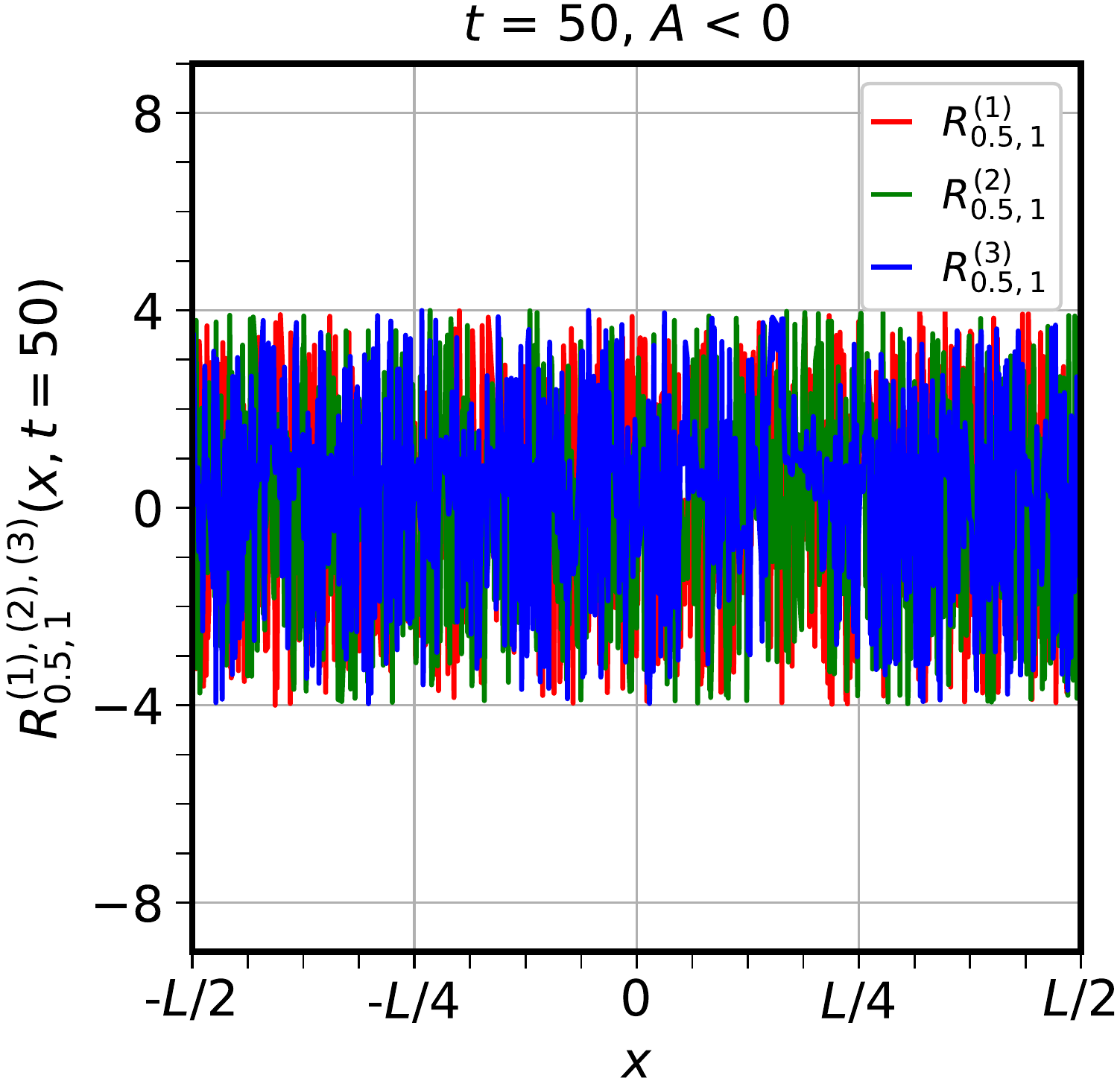}
	\caption{Spatial variation of $E$, $\sigma$, $M_0^2$, $\textbf{M}_1$ in left. Same for $R^{(1, 2, 3)}_{0.5, 1}$ in right. All plots are done at $t = 50$.}
	\label{fig1}
\end{figure*}

\section{Set-up of the system}
Neglecting collisions, vacuum and matter effect the equation of motion in our model governing fast flavor evolution of two flavors of neutrinos moving with velocity $v$ in space ($x$) and time ($t$) is given by \cite{Chakraborty:2016lct}:
\begin{equation}\label{1}
\big(\partial_{t}+v\partial_{x}\big)\textbf{P}_{v}(x, t) = \mu \int_{-1}^{1} dv{'}\left(1-vv{'}\right) \textbf{P}_{{v}{'}}(x, t) \times \textbf{P}_{v}(x, t)\,,
\end{equation}
where the polarization vector $\textbf{P}_{v} = G_v \textbf{S}_{v} = G_v \left(s_v^{(1)}, s_v^{(2)}, s_v^{(3)} \right)^{T}$ represents the flavor composition of neutrinos while $G_v$ and $\mu$ encode the electron lepton number distribution in momentum space and  neutrino self-interaction strength respectively. We present our results for systems with total lepton asymmetry, $\int_{-1}^{1} G_v dv = A = 0$ and $\neq 0$ after solving Eq.\eqref{1} upto the fully nonlinear regime. 
\section{Results}
\textbf{Temporal Behaviour :} In the limit $t$ being very large, our numerical analysis suggests that the fully nonlinear solution of Eq.\eqref{1} becomes \textit{approximately stationary in time} for any $A$. [Fig.\,\ref{fig3} ]

\begin{figure*}[]
	\hspace{-1 cm}
	\includegraphics[width=0.27\columnwidth]{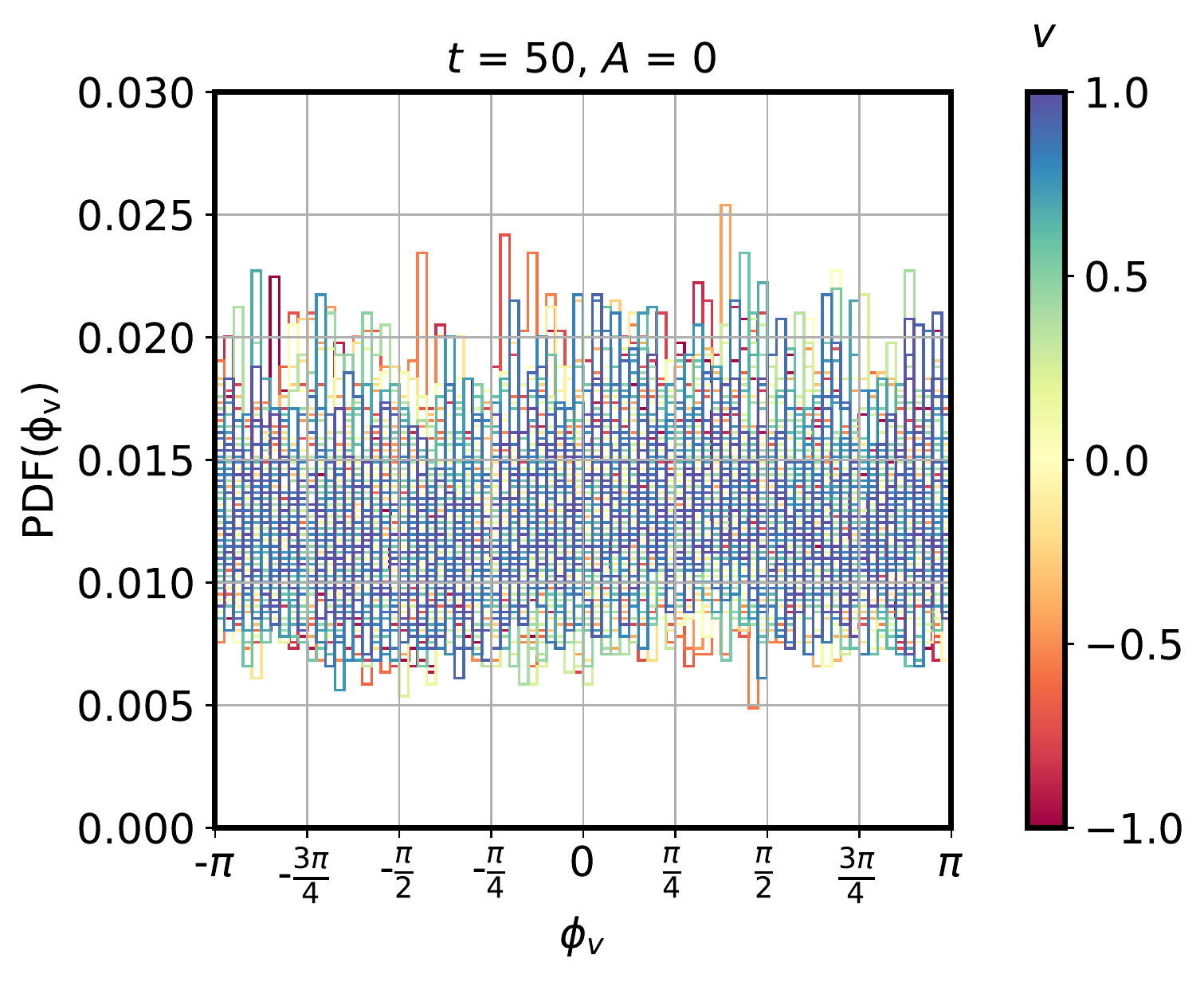}~
	\includegraphics[width=0.27\columnwidth]{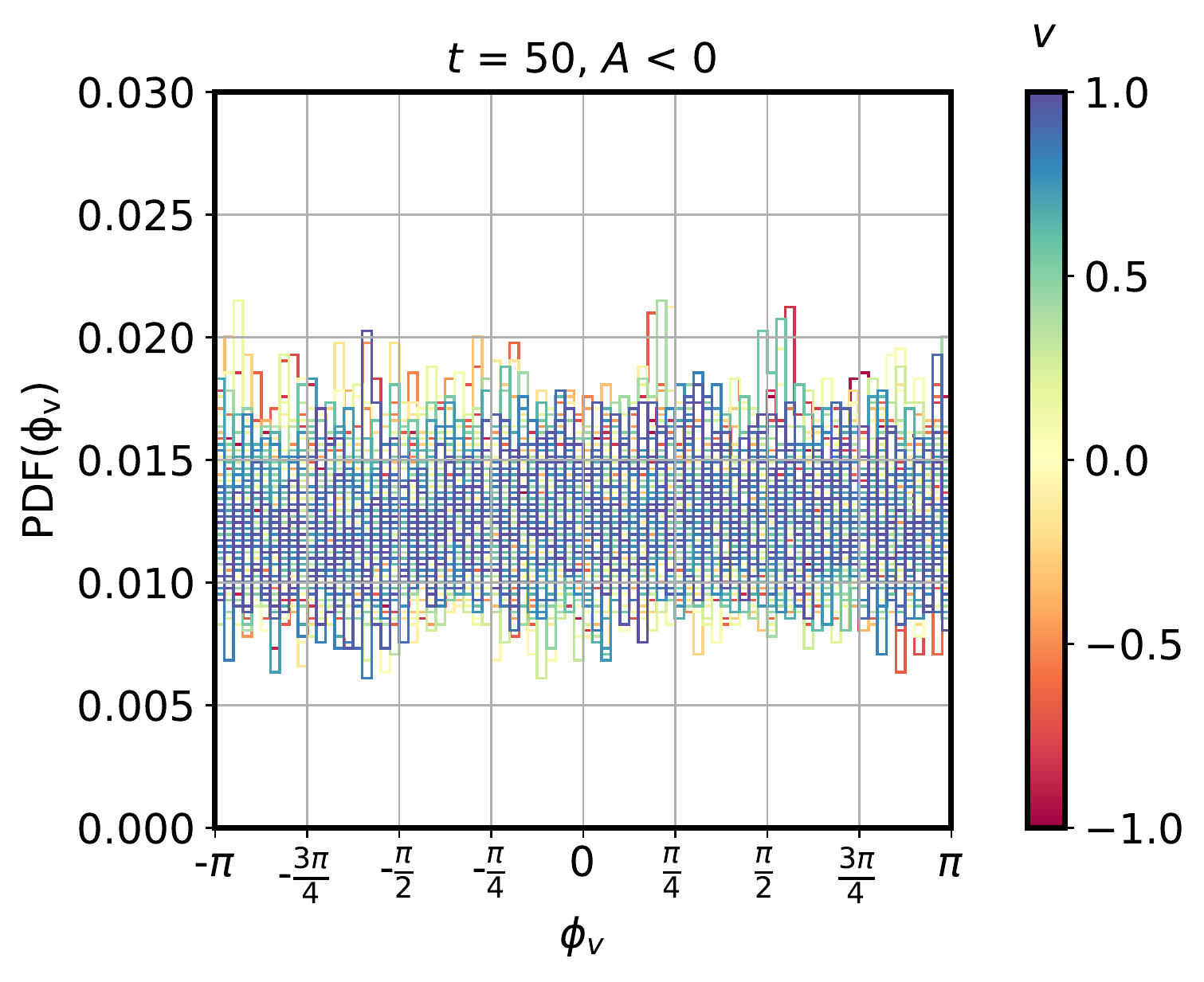}~
	\hspace{0.7 cm}
	\includegraphics[width=0.22\columnwidth]{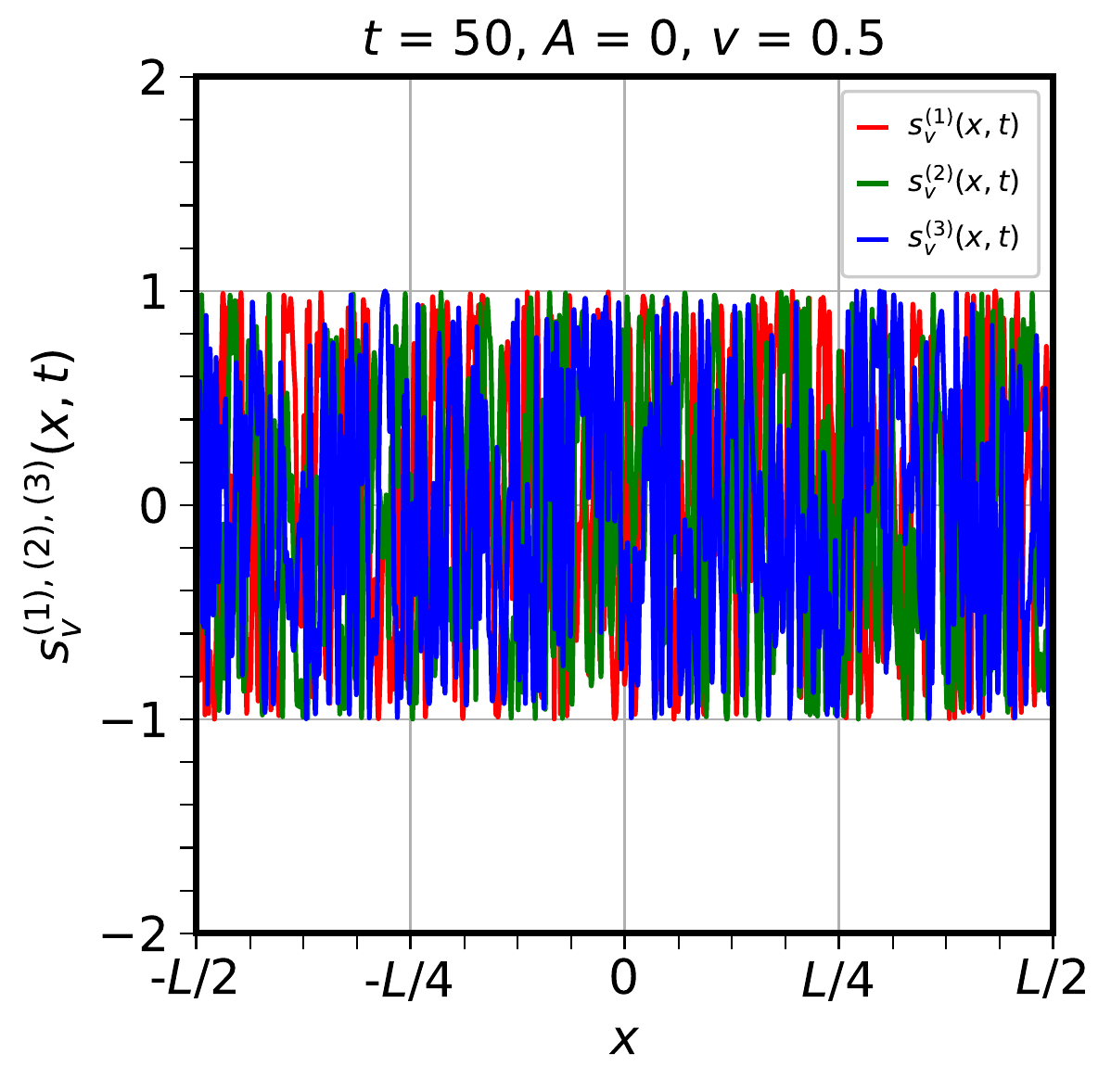}
	\includegraphics[width=0.22\columnwidth]{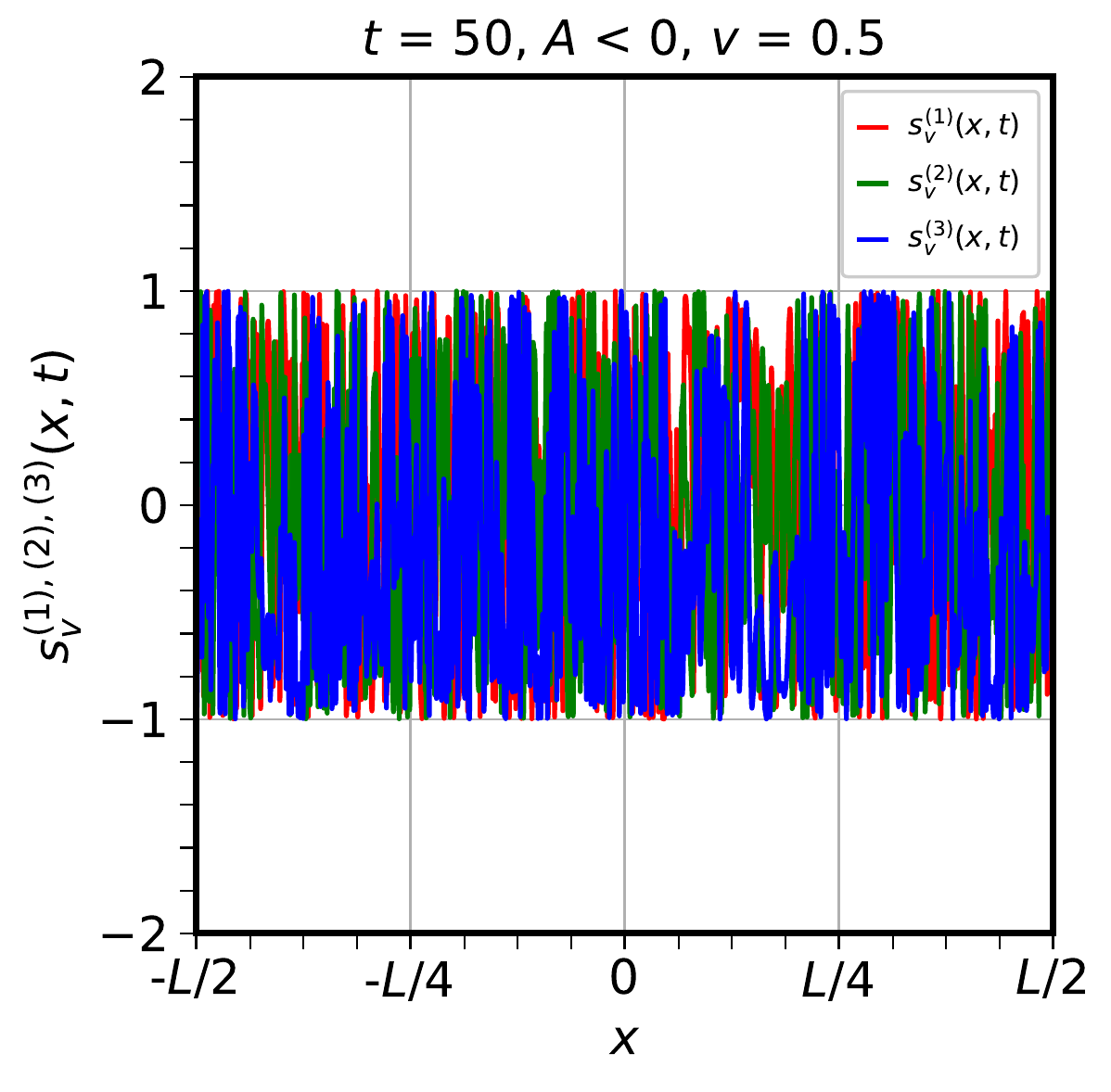}
	\caption{Spatial distribution of $\phi_{v} \big|_{t = 50}$ in left. $s_v^{i} \big|_{t = 50}$ vs $x$ in right shows oscillations indicating precession}
	\label{fig2}
\end{figure*}

\begin{figure*}[]
	\hspace{-0.85 cm}
	\includegraphics[width=0.27\columnwidth]{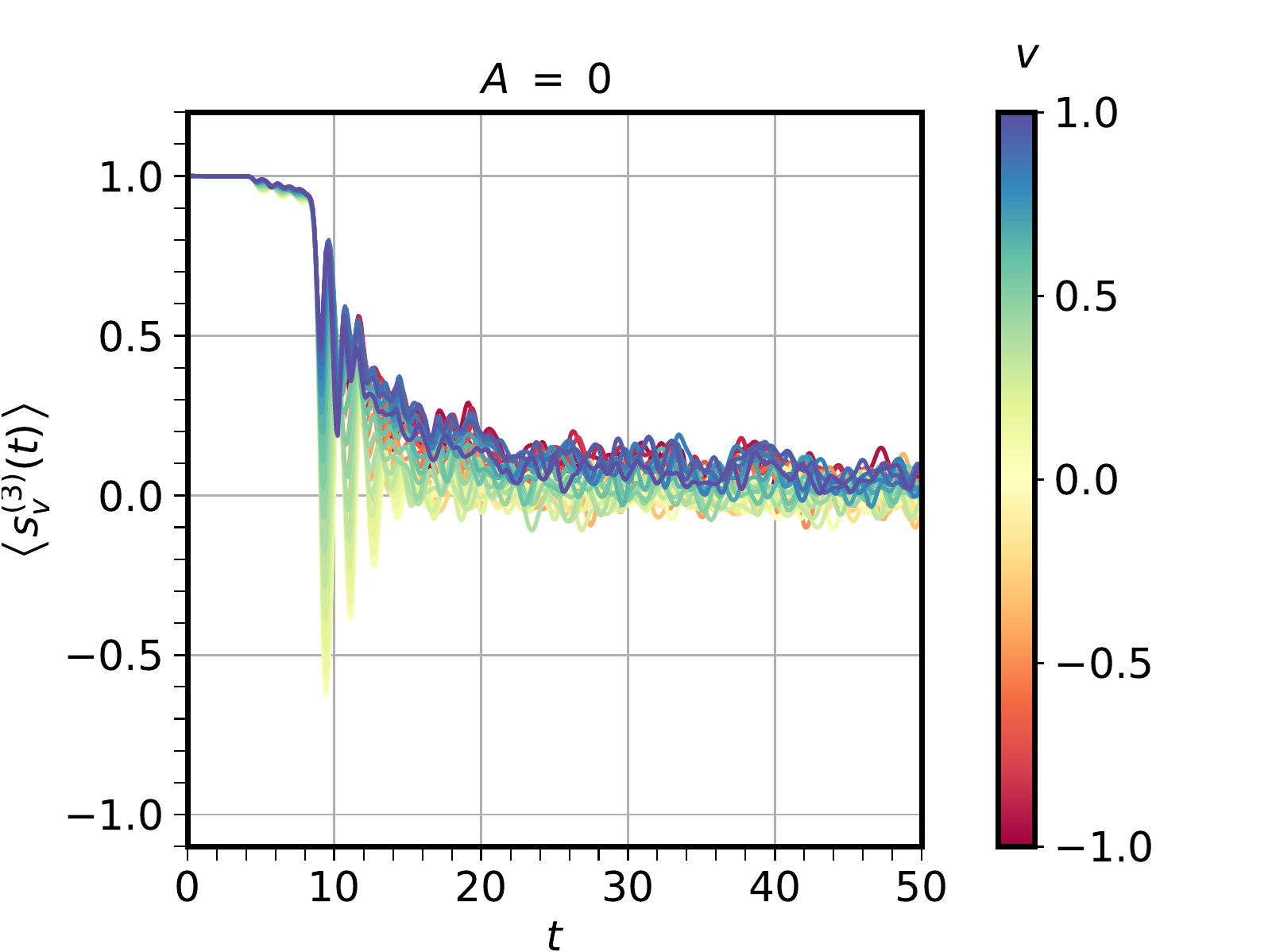}~
	\includegraphics[width=0.27\columnwidth]{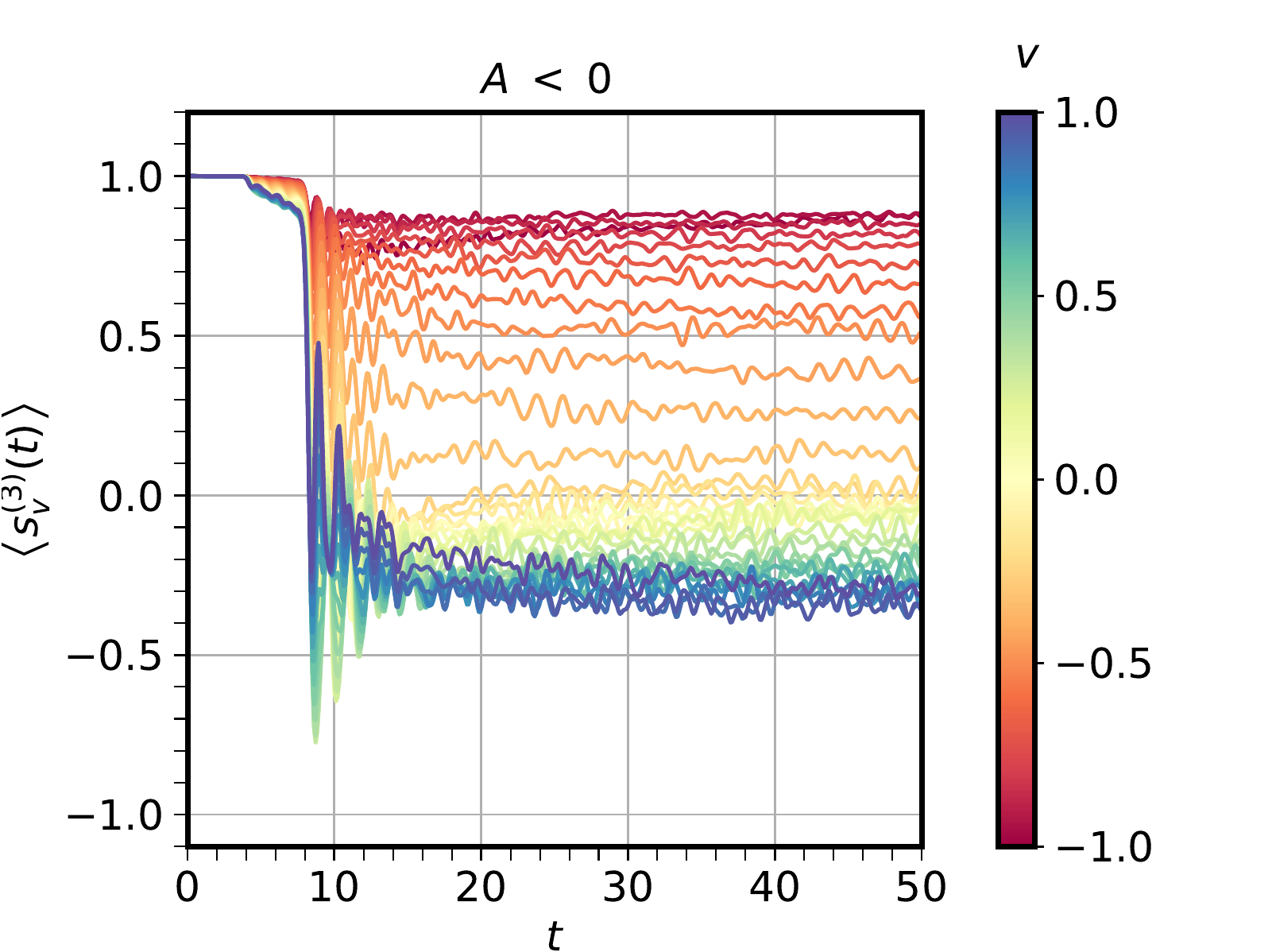}~
	\hspace{0.7 cm}
	\includegraphics[width=0.22\columnwidth]{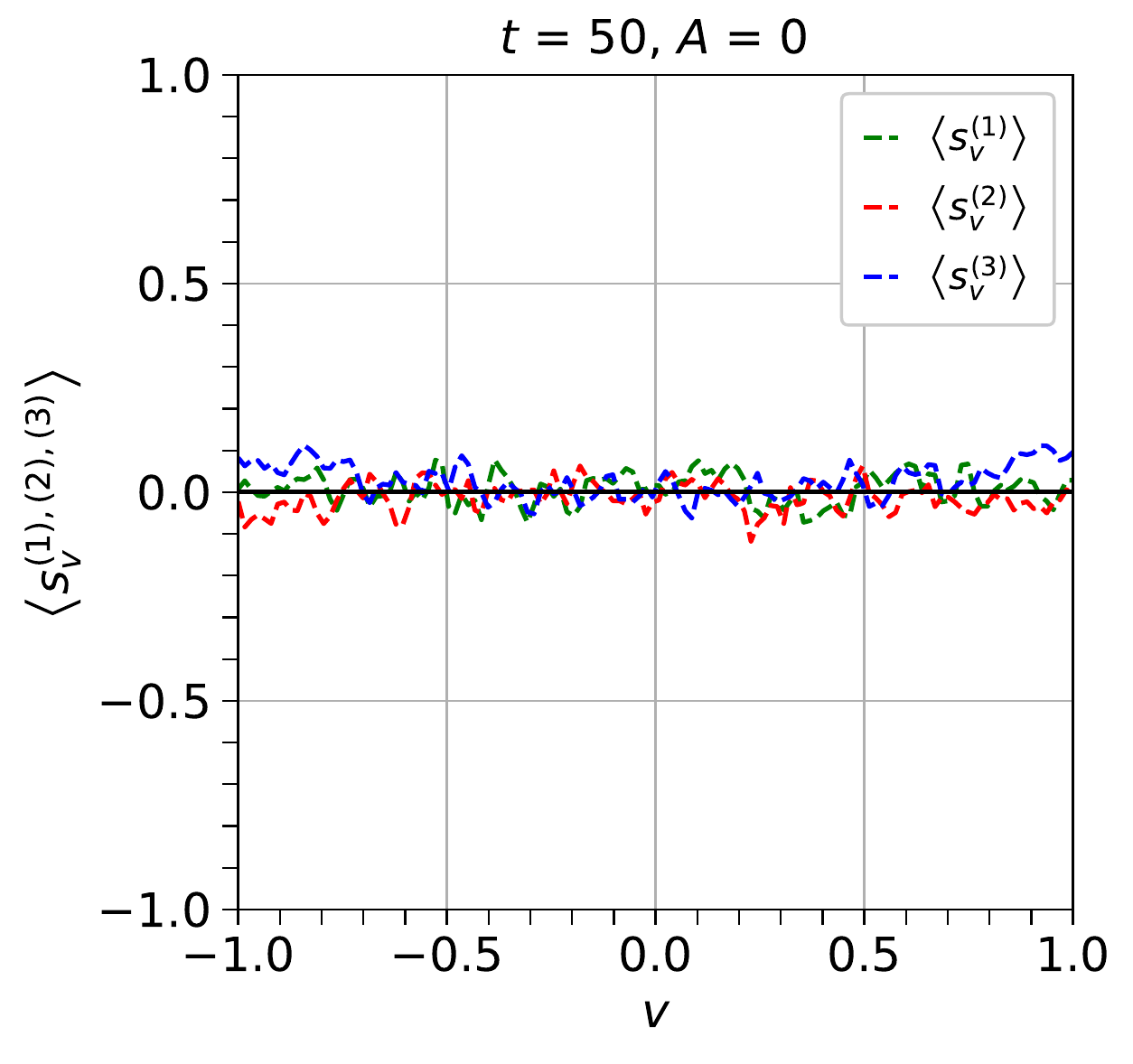}
	\includegraphics[width=0.22\columnwidth]{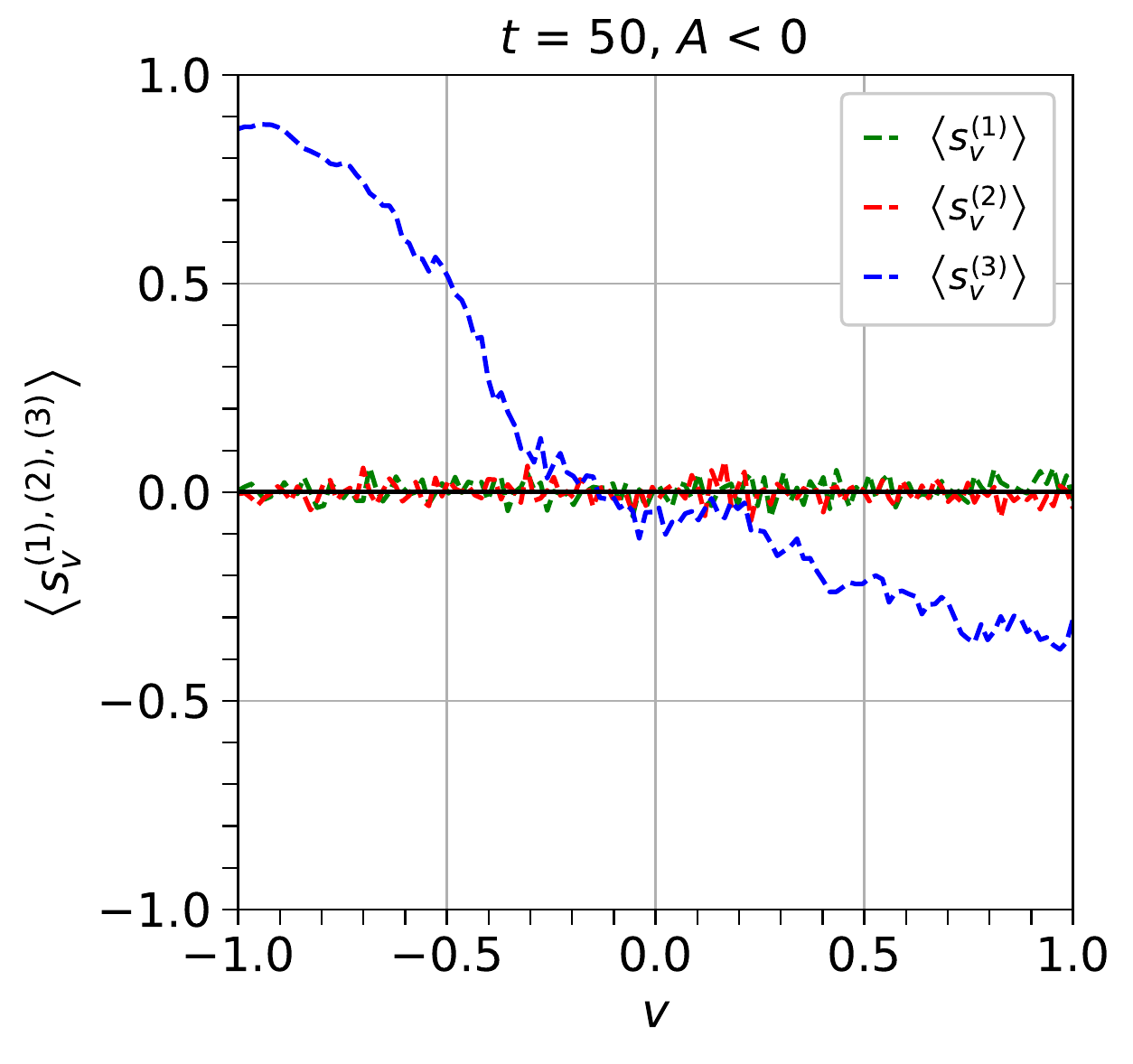}
	\caption{$\langle s_v^{(3)} \rangle$ vs $t$ in left for various $v$ shows steady solution in time. $\langle s_v^{(1, 2, 3)} \rangle$ vs $v$ at $t = 50$ in right.}
	\label{fig3}
\end{figure*}

\textbf{Spatial Behaviour :} From Eq.\eqref{1} using the steady state approximation in time and also going to the multipole space \cite{Johns:2019izj} we obtain the following set of equations governing the system's late time flavor dynamics in a frame rotating with constant velocity $\sqrt{\textbf{M}_1 \cdot \textbf{M}_1}$ w.r.t the fixed lab frame:
\begin{equation}\label{2}
\begin{aligned}
d_{x}\textbf{P}_{v}(x) = \frac{\textbf{M}_{0}(x)}{v} \times \textbf{P}_{v}(x)\,,
\hspace{0.2 cm}
{\textbf{M}}_{0}(x) \times {d}^{\,2}_{x} {\textbf{M}}_{0}(x)+\sigma\,{d}_{x} {\textbf{M}}_{0}(x) =  M_{0}^2  {\textbf{B}}(x)\times {\textbf{M}}_{0}(x) \,, \hspace{0.2 cm} d_{x}\textbf{M}_{1}(x) = 0
\end{aligned}
\end{equation}
In Eq.\eqref{2}, $\textbf{M}_{n} = \int_{-1}^{1}  L_n(v) \textbf{P}_v dv$ denotes the $n^{th}$ multipole moment of $\textbf{P}_v$. This indicates at late times each $\textbf{P}_v$ has a spatial precession of frequency $\frac{1}{v}$ around a common axis, $\textbf{M}_0$, [See Fig.\,\ref{fig2}] which shows \textit{gyroscopic pendulum motion} in space under the action of a \textit{spatially varying magnetic field}, $\textbf{B}(x) = \sum_{r, n=0}^{\infty} c_{rn} {\textbf{M}}_{n}(x)$ maintaining \textit{fixed length} $M_0 = \sqrt{\textbf{M}_0\cdot{\textbf{M}_0}}$, \textit{fixed angular momentum} $\sigma = {\textbf{M}}_{0}\cdot {\textbf{D}}$ and \textit{conserved energy} $E = {\textbf{B}}\cdot {\textbf{M}}_{0}+\frac{1}{2}{\textbf{D}}\cdot {\textbf{D}}$ [See Fig.\,\ref{fig1}]. $c_{rn}$'s are $\left(x, t, v \right)$ independent constants but depend only on the value of $r$ and $n$. Eq.\eqref{2} also indicates a non-separable steady-state solution (non-collective) for $\textbf{P}_{v} (x)$ in position and velocity coordinates \cite{PhysRevD.102.063018}. We numerically checked this via plotting the spatial variation of $R^{(i)}_{v_1, v_2}(x)= \frac{s_{v_1}^{(i)}(x)}{s_{v_2}^{(i)}(x)}$ at $t = 50$ for $i = (1, 2, 3)$ and fixed $\left(v_1, v_2 \right)$ which will be $x$ independent if the solution is collective otherwise not [See Fig.\,\ref{fig1}]. We find the phase relationship ($\phi_v$) between the transverse components of the polarization vector at different spatial locations becomes \textit{randomly distributed} over $[-\pi, \pi]$ at late times for any $A$ and $v$ [Fig.\,\ref{fig2}].

\textbf{Dependence on lepton asymmetry :} Integrating both sides of Eq.\eqref{1} w.r.t all velocity modes and then taking spatial average give rise to three conservation equations: $\int_{-1}^{1} G_v \left\langle \textbf{S}_v \right\rangle = \textbf{A}$ with $\textbf{A} = \left(0, 0, A\right)$. This implies, for $A = 0$ cases the system  can show \textit{synchronized behaviour in momentum space}. But for $A \neq 0$ this is not possible if fast conversions occur and thus system shows a \textit{velocity dependence} at late times to conserve the total lepton asymmetry. [See Fig.\,\ref{fig3}]

\textbf{Momentum dependence :} Naively, the spatial averaged version of Eq.\eqref{1} in some corotating frame looks like, $d_t \left\langle \textbf{S}_{v} \right\rangle = \left\langle \textbf{H}_v \right\rangle \times \left\langle \textbf{S}_{v} \right\rangle$ with $\left\langle \textbf{H}_v \right\rangle = -\frac{\textbf{A}}{3}-v \, \textbf{G}$ and $\textbf{G} = \left\langle \textbf{M}_1 \right\rangle$. This indicates that $\langle s_v^{(3)} \rangle$ for specific $v$ satisfying, $\langle {H}_v^{(3)}(t_{ini}) \rangle \langle {H}_v^{(3)}(t_{fin})\rangle < 0$, or $\left(\frac{A}{3}+v \, G^{(3)}_{ini} \right)\left(\frac{A}{3}+v \, G^{(3)}_{fin} \right) < 0$, \textit{may flip its sign} in the same spirit as \textit{the spectral swaps seen in collective oscillations} \cite{Dasgupta:2009mg}. Clearly, this condition can not be satisfied for $A = 0$ case but in general can be satified for $A \neq 0$ cases. [Fig.\,\ref{fig3}]
\section{Acknowledgments}
S.B would like to thank the organizers of “40th International Conference on High Energy physics- ICHEP2020” for giving an opportunity to present this work as a poster.
\bibliographystyle{JHEP}
\bibliography{references.bib}

\providecommand{\href}[2]{#2}\begingroup\raggedright\begin{thebibliography}{1}

\bibitem{Banerjee:2011fj}
A.~Banerjee, A.~Dighe, and G.~Raffelt, {\it {Linearized flavor-stability
  analysis of dense neutrino streams}},  {\em Phys. Rev. D} {\bf 84} (2011)
  053013, [\href{http://www.arxiv.org/abs/1107.2308}{{\tt 1107.2308}}].

\bibitem{Chakraborty:2016lct}
S.~Chakraborty, R.~S. Hansen, I.~Izaguirre, and G.~Raffelt, {\it {Self-induced
  neutrino flavor conversion without flavor mixing}},  {\em JCAP} {\bf 03}
  (2016) 042, [\href{http://www.arxiv.org/abs/1602.00698}{{\tt 1602.00698}}].

\bibitem{Johns:2019izj}
L.~Johns, H.~Nagakura, G.~M. Fuller, and A.~Burrows, {\it {Neutrino
  oscillations in supernovae: angular moments and fast instabilities}},  {\em
  Phys. Rev. D} {\bf 101} (2020), no.~4 043009,
  [\href{http://www.arxiv.org/abs/1910.05682}{{\tt 1910.05682}}].

\bibitem{PhysRevD.102.063018}
S.~Bhattacharyya and B.~Dasgupta, {\it Late-time behavior of fast neutrino
  oscillations},  {\em Phys. Rev. D} {\bf 102} (Sep, 2020) 063018.

\bibitem{Dasgupta:2009mg}
B.~Dasgupta, A.~Dighe, G.~G. Raffelt, and A.~Y. Smirnov, {\it {Multiple
  Spectral Splits of Supernova Neutrinos}},  {\em Phys. Rev. Lett.} {\bf 103}
  (2009) 051105, [\href{http://www.arxiv.org/abs/0904.3542}{{\tt 0904.3542}}].

\end{thebibliography}\endgroup


\begin{thebibliography}{99}
	 \bibitem{1} S.Bhattacharyya and B.Dasgupta, ``\textit{Late-time behaviour of fast neutrino oscillations}," arXiv:2005.00459 
	 \bibitem{2} S.Hannestad, G.Raffelt, G.Sigl, and Y.Y.Y.Wong, ``\textit{Self-induced conversion in dense
	 	neutrino gases: Pendulum in flavour space}", arXiv:astro-ph/0608695
	 \bibitem{3}S.Chakraborty, R.S.Hansen, I.Izaguirre, G.Raffelt,``\textit{Self-induced neutrino flavor conversion without flavor mixing}", arXiv:1602.00698
\end{thebibliography}
\end{document}